\documentclass[aps,pra,twocolumn,preprintnumbers,amsmath,amssymb,footinbib]{revtex4-1}

\usepackage{listings}
\usepackage{xcolor}
\usepackage{hyperref}
\usepackage{graphicx}
\usepackage{graphicx}
\usepackage{amssymb}
\usepackage{color}

\begin{document}

\title{SwarMS: Swarm-Computations for Mobile Scenarios}

\author{Christopher Scherb}
\email{christopher.scherb@unibas.ch}
\affiliation{University of Basel, Basel, Switzerland}

\author{Pascal Bürklin}
\email{pascal.buerklin@unibas.ch}
\affiliation{University of Basel, Basel, Switzerland}

\author{Christian Tschudin}
\email{christian.tschudin@unibas.ch}
\affiliation{University of Basel, Basel, Switzerland}


\begin{abstract}
Nowadays, most network systems are based on fixed and reliable infrastructure. 
In this context Information Centric Networking (ICN) is a novel network approach,
where data is in the focus instead of hosts. Therefore, requests for data are independent from the
location where the data is actually stored on. 
This property is optimal for infrastructure-less and swarm networking. 
However, the ICN does not provide support for networks without infrastructure.
In this paper we present SwarMS, an architecture for swarm on-boarding, swarm coordination and swarm computations in ICN style networks. 
We use append-only logs to solve the challenges of loose mobile swarm organisation and executing computations. By replicating tasks on multiple nodes we achieve more reliability and trust in results. 
\end{abstract}

\maketitle

\section{Introduction}

In many mobile scenarios coordination between participants is required to complete tasks more efficiently.
Thereby, mobile devices need to organize themselves to form swarms (cooperative units). Within these swarms, tasks can be distributed and each device takes a specific role. This way, all nodes can save resources by completing the full task together and only performing a part of the task themselves.
Data can be shared between swarms in a way that multiple swarms can form a super-swarm.

This kind of mobile infrastructure-less computations have a board range of possible use cases. Starting with simple scenarios where mobile phone user shared data in absence of cellular networks or mobile phone users share an uplink to reduce the load on a base station, there are also complex scenarios such as vehicular networking, drone swarm coordination, or cyber-physical systems.

In simple scenarios, the main task is the coordination of the network and to define routes. For example for infrastructure-less data exchange, there is the requirement that nodes forward data over multiple hops to their destination, while when sharing a cellular connection, one mobile device needs to collect requests and distribute replies over multiple hops.

In more complex scenarios, there is beside of the coordination of the network also a further state synchronization, such as distributed computations, where each node takes a specialized role. In general, roles should be taken by several nodes, so that in case a node leaves the swarm, the swarm can still process the final result. Moreover, a swarm should replace computations former nodes, which left the swarm to achieve and maintain full redundancy. 

The idea of swarm-based computation focuses on highly mobile scenarios, however, each mobile scenario can be extended with stationary nodes. Thus, it is feasible to use the presented technology also in infrastructure-rich scenarios.

In this paper, we present SwarMS -- a mechanism for swarm organization and coordination to cooperatively perform computations. Moreover, we enable swarms to exchange data with other swarms to perform operations as super swarms. 

Thereby, for the coordination as well as for the communication we use synchronized append-only logs and a publish subscribe model.
For our append-only logs, we propose a garbage collection mechanism, which enables us to control the required storage space. 
In the end, we discuss the possibility of a coin based payment system for swarm computations. 

\section{Related Work}
In the past there were proposed several solutions for swarm based communication which use append-only logs and log replication. This work extends them with the capability to perform computations. 

\subsection{Append-Only Logs}
\emph{Append-only logs}~\cite{chun2007attested} are a technology mainly known from block-chains~\cite{swan2015blockchain}. 
An append-only log is a chain of entries, where all entries are secured by cryptographic signatures. Moreover, new entries contain cryptographic secured back pointer, in order to verify, that none of the entries was changed afterwards. 

\subsection{Secure Scuttlebutt}
Secure Scuttlebutt (SSB)~\cite{tarr2019secure} is a social media platform, where data is stored locally on the devices of the user. This way, the always online property of other social networks is removed. SSB uses an \emph{append-only log} technology and each user publishes data in their own \emph{append-only log}. 
Data are secured by cryptographic signatures as used in block-chains~\cite{swan2015blockchain}. However, in SSB each user maintains its own log, and followers/friends append new entries of their friends to their own log. Later, they can further share this information. By digitally securing the content itself, only follower can decrypt and access the information.

Using this mechanism, SSB does not require infrastructure, but synchronizes the logs whenever there is any connection available. It could be an infrastructure based internet connection, but it also can be a \emph{peer-2-peer} connection using bluetooth, wifi-direct, or similar.

\subsection{A Broadcast-Only Communication Model Based on Replicated Append-Only Logs}
Beside SSB and social media, log replication can be used as a communication model~\cite{tschudin2019broadcast}. Nodes publish content and push it as broadcasts to other nodes, which cache the content in their logs. By cryptographic signatures and numbering the messages, it is possible to ensure, that a message is broadcast exactly once and in order, since nodes only forward the next possible message regarding its own log. All other messages are delayed. If a message is delayed, the missing messages must be retransmitted.

\subsection{Information Centric Networking}
Information Centric Networking (ICN)~\cite{jacobson2009networking, zhang2014named} is a novel approach to computer networking, putting the focus on data instead of hosts. Data are directly addressed by interest messages as requests, while the reply data are stored in content objects. Content objects are identified by hierarchical names. 
The pull-based hop by hop communication model enables ICN to perform efficient multicasts and to cache popular data close to the location of potential users. 
Within ICN, there are approaches for mobile and swarm environments, which mainly focus on routing but still keep the basic ICN architecture~\cite{wang2014netwrap, descamps2019swarm}.

\subsection{Named Function Networking}
Named Function Networking (NFN)~\cite{sifalakis2014information, scherb2016packet} is an approach to execute computations within an Information Centric Network (ICN), without specifying where the result is executed. The network automatically finds the best execution location~\cite{scherb2018smart, scherb2019execution}.   
Computations are encoded in interest messages and function code is stored in content objects so that they can be easily forwarded and transferred over ICN.
Important is, that the workflow description which is used to encode the computation in the name of the interest is functional, so that individual components are independent of each other. For example, the $\lambda$-calculus is a possible way to describe computations in interest messages.
NFN supports a network steering operations~\cite{scherb2017execution}, content security~\cite{marxer2016access}, in-network-streaming~\cite{scherb2017network} and with some restrictions executing in mobile environments~\cite{scherb2020resolution}, while an efficient on-boarding mechanism is missing.
Beside NFN there are other in network computation solutions such as SCN~\cite{braun2011service}, CFN~\cite{krol2019compute}, NFaaS~\cite{krol2017nfaas}.

\subsection{Vehicular Communication}
In ICN there are approaches for Vehicular Computing based on the Infrastructure (Vehicle-2-Infrastructure, V2X), based on NFN~\cite{grewe2017information, scherb2018resolution}. Thereby, handovers to other base stations due to the car mobility are handled and results of computations can be delivered over the base station, where the car is connected to, when the result is ready~\cite{scherb2018resolution, grewe2018network, scherb2019data}. 
However, there is a shortcoming in areas where no infrastructure is installed. Common approaches for Vehicle-2-Vehicle (V2V) communication are focusing only on the exchange of information, but not on cooperative computations~\cite{whaiduzzaman2014survey}.

\subsection{UAV Swarm Communication}
Communication in unmanned aircraft systems (UAS) such as drone-swarms breaks up into two different architectures: Infrastructure-based, which relies on supervision of ground control stations (GCSs), that monitor flight operation. Flying Ad-Hoc Networks (FANETs) do not rely on infrastructure, which adds redundancy as well as versatility to the swarm~\cite{8500274}.

\section{A Intra-Swarm-Based Public-Subscribe Model for Computations}
In this section, we present a model of how a swarm can organize itself in a loose way to perform a computation in a cooperative way or to exchange data or results. Thereby, we use a loose coupling of nodes without providing guaranties.
However, we try to minimize the impact of node failures by replication of computations. 
As a communication model, we use an \emph{append-only log} technology, which synchronizes the \emph{append-only logs} between nodes in the swarm.  
In general, each node can simultaneously participate in multiple swarms, by maintaining independent \emph{append-only logs} per swarm.

\subsection{Overview}
In SwarMS we have nodes which form a swarm to cooperatively solve a task.
Nodes inform other nodes by beacons about their existence. 
A node joins a swarm by replicating a request-log that contains the tasks the swarm wants to solve and the corresponding input data. A node can pick a task from the log and solve it. The result will be written into a result-log. 
A node writes information about which tasks it chooses for solving into the request-log.

To improve reliability, a single task is solved by multiple nodes. This way, in case a node leaves a swarm, the overall result can still be produced. 
Moreover, by handing over a task to multiple nodes, the likelihood of fraud is reduced. Thus, if a node leaves a swarm, other nodes can take over the replication so that all tasks are always solved by multiple nodes. 

The results of individual tasks in the result-log can be used to produce the final result. The result-log can be cached, to share the results later to other nodes or other swarms. 

SwarMS uses a NFN-Workflow-Definition, which is functional. It consists of ``ICN-style'' names, which can consist of multiple hierarchical components:
\begin{verbatim}
    /name/component1/component2/...
\end{verbatim}
Function Code is stored in data objects and can be addressed by their names. 
A function call is defined as:
\begin{verbatim}
    /name/of/function_call (List[<parameter>])
\end{verbatim}
where \verb|<parameter>| is either a name or another function call. 

This way, a task can be either to fetch and deliver some data or a function call which should be executed. 
If a function has multiple sub-function-calls, each sub-computation is considered to be an individual task.

\subsection{Architecture}
A SwarMS node contains different components to organize, communicate and perform computations:
\begin{itemize}
    \item a network interface manager,
    \item a neighbor discovery,
    \item a log-synchronizing component,
    \item a log store -- identified by names,
    \item a NFN-style computation executor.
\end{itemize}

\begin{figure}
    \centering
    \includegraphics[width=0.4\textwidth]{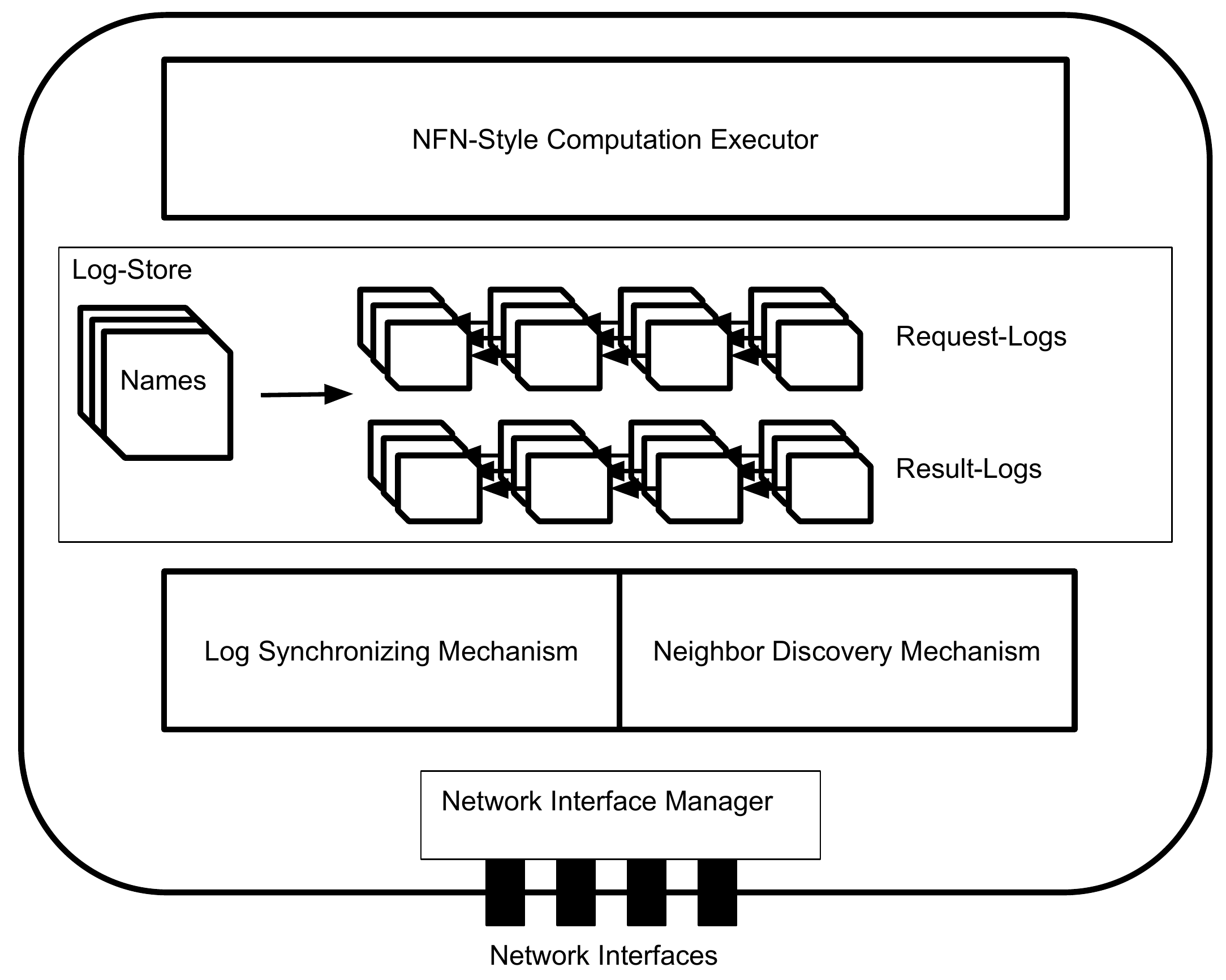}
    \caption{The node architecture of a SwarMS-Node}
    \label{fig:Architecture}
\end{figure}

Figure \ref{fig:Architecture} illustrates the node architecture of a SwarMS node.
The log store maintains two logs for the communication within a swarm: 
\begin{itemize}
    \item a \emph{request-log} and 
    \item a \emph{result-log}.
\end{itemize}

Since a node can participate in multiple swarms simultaneously, for each swarm such a log-pair (request and result log) is maintained.
Each log-pair is identified by a unique identity, using a hierarchical, ICN-style name. In the \emph{request-log} a node writes all requests -- including required input data and function code -- it wants to share with others and where it hopes for cooperation. In the \emph{result-log} all reply messages are stored. After receiving and ``consuming'' (i.e. using) all results, a node can either keep the \emph{result-log} to share it later with other nodes or drop it. The request log only needs to be maintained, as long as there are open requests. 

A request can be formed using an ``ICN'' style name, if data should be requested or as a ``NFN'' style name, if a result should be requested. In the case of NFN, input data can be written into the \emph{request-log}, so that they are shared with the other participants of a swarm. Replied data or intermediate results are being written into the \emph{result-log}.

\subsection{Neighbor Identification}
To identify neighbors, periodical beacons are used. A beacon contains a list of all logs which a node has stored locally. These logs are identified by their name.
Receiving such a beacon, a node can decide, if it wants to replicate a specific log. 
To replicate a log, a node will answer to a beacon, to inform the sender to transfer a specific log. The answer to a beacon will contain all names of logs that the node wants to replicate. 
By replicating a log recursively, the node becomes part of a swarm and also enables all nodes in range to replicate the log and to join the swarm.
Thus, a swarm does not only grow if more nodes are joining and replicating the swarm's logs, but it also becomes more prominent. 

A node identifies itself with a cryptographic signature. Next, the number of available logs on the node is stored beside a field that flags, if all available log-names are stored within the beacon or if there are more names, that does not fit in the frame. Last, the names of the first $N$ logs are stored in the beacon.
If not all nodes are stored in the beacon, a neighbor can react to the beacon and request the names of the remaining logs. 

\subsection{Log Synchronization}
If a node replies to a neighbor discovery beacon, the log synchronizing mechanism kicks in. 
The log synchronization subscribes to all updates to the chosen log (both the request- and the result-log).
Thus, whenever a node appends a new message to its own copy of the log, all nodes replicate the new entry. This way the status of the swarm is synchronized. 
In case, multiple append events happen simultaneously and lead to a different order on different nodes, we sort the data always first by timestamp and second by the private key of the publishing node. This means, for some nodes, in very rare cases it might be necessary to recompute some of the check sums to restore the order of the messages. This needs to be done by the content producer. However, this principle keeps very easy to verify no node changed any data by verifying the last signatures on all nodes. 
Since in NFN individual parts of a computation are independent of each other due to the functional manner maintaining a more specific order is not necessary. 

Each log component is individually signed by its creator, so that nodes cannot change the content without breaking the signature. 

In general, the log synchronization mechanism starts by synchronizing the request log. Afterwards, it synchronizes the result log. 
Now, a node can verify, which tasks are already solved by comparing the request with the result log. Next, the node can pick a message from the request log. When it does this, it appends a note in the request log, that it has taken the task.
Other nodes, checking the logs can use this information to choose another computation, or replicate already running computations as trust verification.

When a node picks a computation, from the request log, it should not choose the first computation which is not already taken or not enough replicated but a random one, since the log synchronisation may take some time. 

\subsection{The Logs and the Log-Store}
The log-store is the important data structure, which maintains the logs stored on a node and by this it also maintains the swarm membership. 
Each entry in the log-store is identified by an ``ICN-style'' name. Moreover, each entry contains a request-log and a result-log. 

A log can contain different data types, depending on how it is used and depending on the state of the computation in the swarm. The main data types are:
\begin{itemize}
    \item Requests: a node describes what data or result it wants to request.
    \item Data: an actual data object containing input data or result information.
    \item Take-Over note: a node writes a message with this data type, to inform other nodes, that it is taking over a request.
    \item Keep-Alive beacon: a message nodes periodically write into the request log to notify other nodes, that they are still working on a task. 
\end{itemize}

In general, only data -- more specific: result data -- can be written into the result log. All other messages will be written into the request log. 
After a request is complete, the request-log can be purged, since maintaining input data and requests are not required anymore. However, the result log, which contains the final results as well as sub-results can be kept for some time, to share information with nodes joining the swarm late or with other swarms. 

\begin{figure*}
    \centering
    \includegraphics[width=0.7\textwidth]{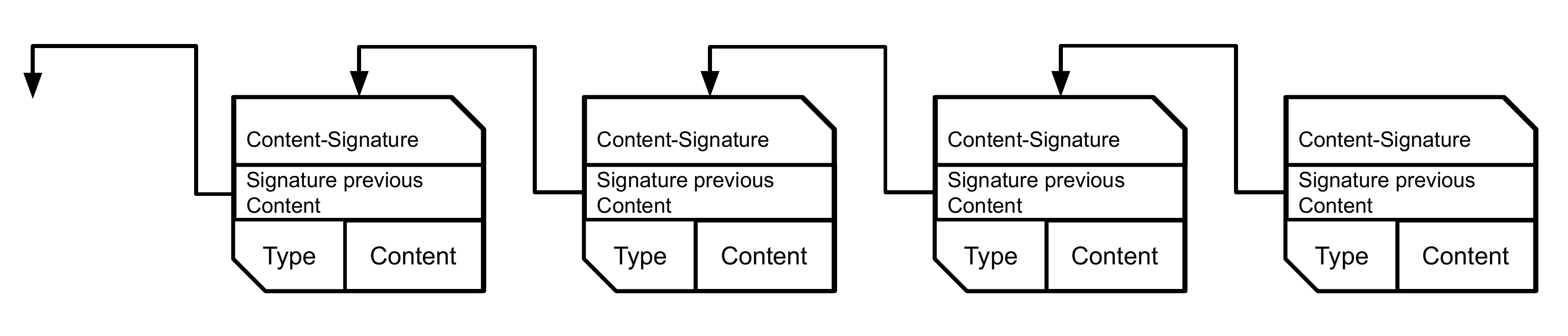}
    \caption{The log architecture used by SwarMS.}
    \label{fig:Log_Architecture}
\end{figure*}

Our log architecture is designed as shown in Figure \ref{fig:Log_Architecture}.
Each content itself is signed by the key of the producer and contains a back-pointer to the previous content in the log.
The back-pointer is part of the signed content, which ensures that previous entries cannot be changed by single nodes anymore. The content itself is stored after the back-pointer and contains two fields. First the type of the content and second is the content itself. The type is either request, data, take-over note or keep-alive beacon. 

The first entry in each log is provided as an anchor to be able to verify the signatures of the whole chain. This entry is provided by the node issuing the request and starting to form the swarm. 

If a node wants to verify a result, it needs to verify the log-chain it received and request the last log-entry from different nodes. If these match, the log was not modified by a single node. 

\subsection{Keep-Alive-Beacons}

In case a node leaves a swarm, the number of replications is decreased. Therefore, other nodes should take over the tasks to restore the previous replication status.
Therefore, it is required to detect, if a node left the swarm. 
In SwarMS, this is done by Keep-Alive-Beacons. A node which took over a task periodically writes a message into the request log, that it is still working on the task, If this message was multiple times not found in the log replicated by the other nodes, other nodes take over the task after a random chosen delay. This delay is required to synchronise the new take-over note, so that not all nodes start taking over this task.  

\section{An Inter-Swarm-Based Data Exchange Model}

Up to this point, we described how to join a swarm in SwarMS and how a swarm cooperates to solve a task. 

In the following we will take a look into inter-swarm communication and inter swarm data exchange. 
After a result has been produced, the result log can be maintained to share data with other nodes which were not part of the swarm. 

In general, we use a log cache, which adds a Time-To-Live (TTL) value to each log-pair. After the TTL value the log will be pruned. In case the log was replicated by another node, the TTL value is reset and doubled. This way, popular results stay longer in the network and the probability of reusing results is increased. 

Moreover, depending on the task, a longer or shorter TTL value can be set in the beginning. If the node already knows, the result is only meaningful for a short time (for example, data about number of vehicles on a road can change quickly) long TTL does not make sense. 

In SwarMS, results are stored in a chain, this way, results in the log cannot be easily changed without breaking the signature-chain. Therefore, other nodes, which replicate the log can verify that no changes happened. Since some of the entries in the log are signed by different nodes, a single node cannot easily manipulate the results. By replication the last entries in the log from a second node, the integrity of the data can be further ensured.







\section{A Coin Based Ecosystem for computations}
SwarMS assume nodes want to cooperate to perform a task. However, even if that might be true for some scenarios like vehicular computing (vehicles that are cooperating, can increase their own safety), in many cases helping others to solve a task creates costs without benefit for the own node. 

Therefore, we propose a Coin based Ecosystem, which rewards nodes for participating in a swarm and helping to solve tasks. 
Nodes earn coins and can use these coins to pay for computation results themselves.

The idea behind the coins is very simple:
Each node starts with a certain number $c$ of coins. Whenever, a node receives help from other nodes, it needs to pay coins for the result. Whenever, a node provides a result for someone else, it will receive coins. 

The amount of coins owned by a user can be stored in a block-chain~\cite{nakamoto2019bitcoin}, so we have a distributed storage of coins. In such a system, coins cannot only be earned by performing computations but also by buying them over an external platform coin market.  
This way, if a node wants to benefit from SwarMS, either it needs to participate in a swarm to earn coins or it needs to actually buy coins. 

To ensure payments in a distributed environment, smart contracts~\cite{buterin2014next} can be used. The smart contract is fulfilled, when the result is delivered and thus the coins are handed over.
\vspace{-0.3cm}
\subsection{The Law of Supply and Demand}
The price of computations can vary on the complexity. In SwarMS, we want the price to be regulated by the law of supply and demand. 
A node $n_1$ offers a certain number of coins $c_n1$ per task $t_1$ which it wants to be offloaded to a swarm. If a node $n_2$ joins the swarm, it sees the number of coins, which is offered to perform the task $t_1$. If $n_2$ accepts, it will be rewarded by the coins when it delivers the result. 
If the node $n_1$ does not find nodes that want to take over tasks, it needs to improve the offer. If the node $n_1$ finds nodes that take over the task, for the next tasks it will offer slightly less coins. 
This way, the price for computation can vary. Moreover, for important tasks, by offering more coins, a node can achieve a higher level of redundancy and thus more trustful results. 

\subsection{Pricing for Cached Results}
Previously, our result-log was maintained using a TTL after all task were completed. The caching of the result-logs can be integrated into the Coin based Ecosystem. 
Instead of adding a TTL value to the result-log, we add a price. A node can decide itself about how to fix the first price. It depends on the number of coins it needed to pay to get the result. The price for the cached result-log can be the price the node payed or half or a quarter of it or any value the node decides. 
However, if nobody is paying for the cached result, the node needs to reduce the price for the result. If many other nodes copy the result, the price can be increased. If the price for the cached results hits zero or falls below a defined threshold the result-log will be pruned.

\subsection{Keeping the Equilibrium}
In our Coin based Ecosystem there should be an equilibrium, which ensures that results stay effortable and nodes can afford them.
A problem could be, that when nodes increase the prices for popular results, the price could be rising to a level, that some nodes cannot afford certain results anymore and so they would drop out of the ecosystem. Other nodes, which already have such a popular result can easily distribute it, since data can be copied to almost zero costs.
Therefore, the price of a result needs to be capped. 
To introduce a cap, every node checks the price to compute a result locally by itself. When the node gets in the situation, that it does not find another node which wants to take over a computation, it increases the price. But only to a level slightly below the cost to produce the whole result locally. If the network price exceeds the local price, the node will perform the computation locally. 
This way, no node will pay an excessive price for a result or a cached result and the maximum price for a result is capped on the level of the local production costs and nodes can benefit from cooperation and caching by paying cheaper prices, but do not end up in paying a higher price then necessary.  
\vspace{-0.2cm}
\section{Discussion}
In general, SwarMS is focused on environments, where no infrastructure is available and nodes need to organize themselves using peer-2-peer connections. 
However, SwarMS is not limited to infrastructure less scenarios. As described in the paper, by adding new synchronized logs to the neighbor identification beacons a swarm can exist over multiple hops. Thus, it is also possible to add infrastructure nodes to a swarm, which support computations. More over, synchronizing logs can be used as infrastructure communication. Assuming infrastructure nodes synchronize the request-logs till they find requested data in a data center and then synchronize back the result logs we have a pull based communication over multiple hops in the infrastructure. And in fact, the network behavior becomes very similar to the behavior of ICN (e.g. Named Data Networking, NDN). 
A already synchronized request-log does not need to be synchronised a second time, if another client-node wants to get the same data. This replicates the behavior of the Pending Interest Table (PIT). By only letting those infrastructure-nodes replicate the request-log, which know that they are responsible for this kind of request, we can replicate the Forwarding Information Base (FIB). In absence of this knowledge any infrastructure-node can replicate the log, thus we have a controlled broadcast. 
When the requested data is delivered, the result-log is kept for a certain TTL while the request-log is purged. Thus, the request-log behaves very similar to the PIT while the result-log behaves very similar to the Content Store (CS).
By exchanging the requests from ``ICN-style'' to ``NFN-style'', we also can get the behavior of NFN. Therefore it is required that only infrastructure-nodes which have the input data stored locally start executing the tasks. 
Therefore SwarMS can be seen as ICN/NFN implementation for infrastructure-less and swarm environments, while when used on infrastructure it replicates the classic ICN/NFN behavior.

In SwarMS we have a clear separation between the request-log and the result-log. While the result-log contains only result data objects, the request log can contain also requests, take-over notes and keep-alive beacons. Input data are stored in the request-log. One may wonder, why we choose a request-log with mixed data types and not a single log per datatype. The reason for this is, that we concluded that with regard to reusing of results it is more useful to separate between data which are required to perform the tasks (request-log) and resulting data (result-log). 

\vspace{-0.2cm}

\section{Conclusion}
In this paper we present SwarMS, a swarm based computation system. This paper
presents the architecture and the basic execution principles. 
SwaMS points on environments, where no infrastructure is available and nodes need to organize themselves to a swarm.
A swarm is a loose coupling between nodes to cooperatively perform a task, which can be split into different sub-tasks. 
Possible scenarios are vehicular networking, drone swarms or cyberphysical systems. 
Nodes will use distributed append-only logs to distribute the tasks and to collect results. Using an ``ICN'' style naming and ``NFN'' expressions to describe computations, SwarMS results can be easily cached and reused.
SwarMS is designed for infrastructure-less environments, but by synchronizing log over multiple nodes it can be used in multi-hop environment or even in scenarios where infrastructure is available.

\bibliographystyle{ACM-Reference-Format}
\bibliography{bib/ms}

\end{document}